\documentclass[aps,pra,twocolumn,showpacs,groupedaddress]{revtex4}

\usepackage{graphicx}
\usepackage{amsthm}
\usepackage[toc,page,title,titletoc,header]{appendix}
\usepackage{graphicx}
\usepackage{epstopdf}
\usepackage{CJK}
\theoremstyle{plain}
\newtheorem{thm}{Theorem}

\theoremstyle{definition}

\theoremstyle{remark}

\newcommand{\ket}[1]{|#1\rangle}
\newcommand{\bra}[1]{\langle#1|}
\begin{document}
\begin{CJK*}{GB}{gbsn}
\title{ Engineering of Quantum State by Time-Dependent Decoherence-Free Subspaces}

\author{S. L. Wu(ÎäËÉÁÖ)}

\email{slwu@dlnu.edu.cn}

\affiliation{School of Physics and Materials Engineering,\\
Dalian Nationalities University, Dalian 116600 China}

\begin{abstract}
We apply the time-dependent decoherence-free subspace theory to a Markovian open quantum system in order to present a novel proposal for quantum-state engineering program. By quantifying the purity of the quantum state, we verify that the quantum-state engineering process designed via our method is completely unitary within any total engineering time. Even though the controls on the open quantum system are not perfect, the asymptotic purity is still robust. Owing to its ability to completely resist decoherence and the lack of restraint in terms of the total engineering time, our proposal is suitable for multitask quantum-state engineering program. Therefore, this proposal is not only useful for achieving the quantum-state engineering program experimentally, it also helps us build both a quantum simulation and quantum information equipment in reality.
\end{abstract}

 \maketitle

\end{CJK*}

\section{Introduction}

Controlled manipulation by atoms and molecules using external controls, known as quantum-state engineering (QSE), has become an active field of modern research, which is a fundamental step in quantum computation \cite{qc} and quantum measurement tasks \cite{qo}. The adiabatic theorem of quantum mechanics provides a reliable method of controlling the quantum state of an isolated system \cite{adi1,adi2}. Indeed, for the scheme of QSE, there are two unique advantages for the adiabatic method. Firstly, the adiabatic method of QSE is robust when there is fluctuation in the coherent control fields. Secondly, since the parameters in Hamiltonian varies adiabatically, the engineering timing does not need to be strictly controlled in order to be manipulated precisely. If the QSE process is accomplished, the quantum state will be steadied on the target state. Because of the advantages mentioned above, the adiabatic method has been chosen as an important part of the QSE program and experimentally realized through a number of techniques, such as nuclear magnetic resonance (NMR) \cite{nmr1,nmr2}, superconducting qubits \cite{sc}, trapped ions \cite{ti}, and optical lattices \cite{ol}.

When the quantum system is coupled to its surroundings, the adiabatic QSE process will experience considerable loss of fidelity, which limits the application of the adiabatic method. Actually, for open quantum systems, there is competition between the time required for adiabaticity and the decoherence time scales \cite{comp1,comp2}. Therefore, identifying a protocol that is both fast and fault-tolerant is an important research direction for quantum information processing and quantum control. Many enlightening proposals have also been put forward and evaluated for non-adiabatically engineered quantum states, such as the inverse engineering control \cite{nadi1}, optimal control\cite{nadi2}, the fast quench dynamics method \cite{nadi3}, and a method by combining incoherent and coherent controls \cite{nadi4,nadi5}. The fundamental idea of these methods is to decrease the time needed to manipulate the quantum state so as to reduce the effect of decoherence on fidelity. Obviously, this is not enough to realize quantum information processes for real applications. On the one hand, the quantum state will lose its quantum characters {(the coherence between two quantum states or the entanglement between two quantum system)} and decay into the steady state over time or through repeated operation on this quantum system. Thus it limits the QSE scheme in terms of achieving a multi-task QSE program. On the other hand, the success of ultra-fast QSE is determined by the fact that control of the quantum system must be ultra-precise and ultra-fast, which strongly depends on the development of experimental technology.

In this paper, we propose a novel method to engineer the quantum state of an open system. With this innovative method, there is no limit to the total engineering time nor any loss of fidelity. Our QSE method was designed based on the time-dependent decoherence-free subspace (t-DFS) scheme \cite{nadi4,tdfs2} in which the basic vectors are time-dependent. In other words, such a DFS evolves smoothly in the total Hilbert Space of the open quantum system by reservoir engineering technology \cite{se1,se2}. If we manipulate the quantum system state properly, the quantum state will strictly follow the evolution of the t-DFS, so as to protect it from the effect of decoherence. In comparison with existing works on QSE, our method is more effective and promising. Because the DFS scheme can act against decoherence completely, the quantum state in the DFS scheme does not lose any quantum character. More importantly, when the quantum state involves following the evolution of the t-DFS strictly, the QSE process is unitary, as if the environment does not exist. Therefore, no matter how much time is spent on the QSE process, the target state will be reached with no loss of fidelity. Owing to the distinguishing features of our method, it offers a reliable path to implementing a multi-task QSE process on identical open quantum system one after another. Moreover, the robust QSE program can also be realized in a time-independent DFS. A DSF of at least two-dimensions is required for the simplest QSE task, which means that we have to control several quantum systems at the same time. For our method, since the basic vectors of t-DFS are time-dependent, a one-dimensional t-DFS is sufficient to accomplish all QSE tasks in principle.

To illustrate the practical application of our method, a QSE program of a two-level open quantum system was designed according to the t-DFS scheme. As shown in the results, the QSE process was completely unitary even over a long period of time. {The analytical expression of the coherent control field was derived, by which we could check the QSE process in detail.} As presented in the analytic expression, there was a singular point in the coherent control field that could not be reached in actual experiments. Therefore, we introduced some adjustments to the coherent control field, and the results remained satisfactory. The QSE process is always robust, even over the long-term. Thus we can affirmatively conclude that our method is powerful and reliable in both its theoretical preciseness and experimental feasibility.

This paper is structured as follows. In Sec. \ref{TDFS}, we briefly review the t-DFS scheme and discuss how to engineer a quantum state within such a t-DFS. In Sec. \ref{SNC}, we manipulate a two-level open system to the target state by means of the t-DFS QSE method. Both population engineering and phase engineering are discussed step by step. An adjustment on the non-physical coherent control field is considered in Sec. \ref{AD} in order to show that even when the coherent control field is defective, the t-DFS  QSE method is still unconditionally robust. We conclude with Sec. \ref{CC}.

\section{The Time-dependent Decoherence-Free Subspaces and the Quantum State Engineering Program}\label{TDFS}

Let us start with the t-DFS scheme. A DFS is a subspace of the Hilbert space of the open quantum system, in which the dynamics of the quantum system are still unitary \cite{dfs1}. It has been shown that the principle behind such a charming appearance of DFS is symmetry of the interaction between the open quantum system and the environment. The existence of DFS has been demonstrated experimentally in many physical systems \cite{dfs2,dfs3,dfs4}, and many enlightening designs have also been proposed based on DFSs in order to realize quantum key distribution \cite{qkd}, quantum computation \cite{qc2}, and so on.

Although DFS with fixed basic vectors (traditional DFS) is a promising candidate for quantum information processes, it is more suitable to storing and protecting the information coded in quantum systems; however, in a QSE field, DSF is unable to manipulate quantum states precisely. For instance, at least three physical qubits are needed to construct one logical qubit against the effect of a dephasing environment, and the quantum computation program on such a logical qubit needs to accurately control the interactions between the physical qubits \cite{qc1}, i.e., two interactions have to be controlled simultaneously. However, it is difficult to manage the couplings between the physical qubits at the same time. Moreover, the decoherence becomes more complicated when the number of physical qubits and energy levels increases.  In order to conquer these difficulties, the t-DFS scheme is introduced \cite{nadi4,tdfs2}.The t-DFS is still a DFS, but its basic vectors depend on time, which means that the t-DFS will evolve in the Hilbert space of the quantum system.

In the following, we restrict our discussion to an $N$-dimensional open quantum system and consider its dynamics as Markovian. In the interaction picture, the evolution of the quantum system must obey the Lindblad--Markovian master equation, given as
\begin{eqnarray}
\dot{\rho}(t)&=&-i[H,\rho(t)]+\mathcal{L}\rho(t),\nonumber\\
\mathcal{L}\rho(t)&=&\sum_\alpha \left[F_\alpha \rho(t)
F_\alpha^\dag-\frac{1}{2}\{F_\alpha^\dag F_\alpha,
\rho(t)\}\right], \label{tmq}
\end{eqnarray}%

where $F_\alpha$ is the Lindblad operator. which describe the decoherence caused by the coupling to the environment, and $H$ is the Hamiltonian, which consists of the coherent control field on the open quantum system. It has been shown in Ref. \cite{tdfs3} that if some of the environment parameters can be continuously varied as a function of time by means of reservoir engineering technology, the Lindblad operators in Eq. (\ref{tmq}) will be time-dependent. In other words, when the environment varies with time, the symmetry of the interaction between the open quantum system and its environment is time-dependent \cite{svo,sva}. It is also a way of engineering the state of the open quantum system, which is known as the incoherent control method.

In the context of the Linblad--Markovian master equation, the DFS is defined as a collection of quantum states in which the dynamics are unitary and the purity is constant as the evolution of the quantum states $\rho(t)$, i.e., $\partial Tr[\rho^2(t)]/\partial t=0$, leading to the following conditions on the t-DFS \cite{tdfs2}:
\begin{thm}
Let the time evolution of an open quantum system in a
finite-dimensional Hilbert space be governed by Eq. (\ref{tmq}) with
time-dependent Hamiltonian $ H(t)$ and time-dependent Lindblad
operators $F_\alpha(t)$. The subspace
\begin{eqnarray}
\mathcal
H_{\text{DFS}}(t)=\text{Span}\{\ket{\Phi_1(t)},\ket{\Phi_2(t)},
\cdot\cdot\cdot,\ket{\Phi_M(t)}\}
\end{eqnarray}%
is a t-DFS if and only if each basis vector of $\mathcal
H_{\text{DFS}}(t)$ satisfies
\begin{eqnarray}
F_\alpha(t)\ket{\Phi_j(t)}=c_\alpha(t)\ket{\Phi_j(t)}, j=1,...,M;
\alpha=1,...,K, \nonumber\\ \label{sandn}
\end{eqnarray}%
and $\mathcal H_{\text{DFS}}(t)$ is invariant under
\begin{eqnarray}
H_{\text{eff}}(t)&=&G(t)+
H(t)\nonumber\\&&+\frac{i}{2}\sum_\alpha\left(c^*_\alpha(t)
F_\alpha(t)-c_\alpha(t) F^\dag_\alpha(t)\right). \label{heff}
\end{eqnarray}%
Here $G(t)=iU^\dag(t)\dot U(t)$  and $U(t)$ is an unitary operator
\begin{eqnarray}
 U(t)=\sum_{j=1}^M
\ket{\Phi_j(0)}\bra{\Phi_j(t)}+\sum_{n=1}^{N-M}
\ket{\Phi_n^\bot(0)}\bra{\Phi_n^\bot(t)}.\label{unitary}
\end{eqnarray}%
\end{thm}
In this theorem, the time-dependent Hamiltonian describe the coherent control on the open quantum system, which can be rewritten as $H(t)=\sum_n \Omega_n(t)H_n$ with the control Hamiltonian $H_n$ and the coherent control field $\Omega_n(t)$.{In our pervious work\cite{tdfs2}, we have proved that the theorem mentioned here is a sufficient and necessary condition for existing the t-DFS. In the following, we will apply this theorem to design our QSE program, and investigate the coherent and incoherent control projects in detail.}

\begin{figure}
\includegraphics[scale=0.45]{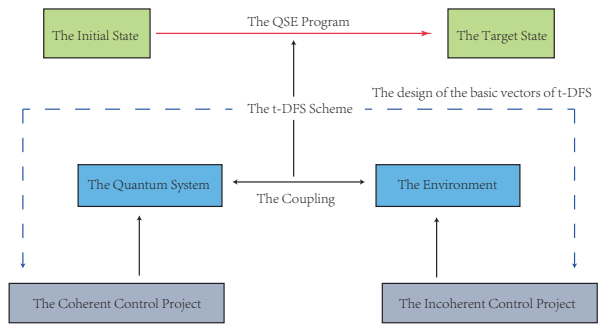}
\caption{(Color online) A schematic diagram for the t-DFS QSE program. The solid lines are the control process of the t-DFS, whereas the dash lines are to illustrate the design principle of the t-DFS. The goal of the QSE program is to design a path in the Hilbert space to connect the initial state with the target state, which is assisted by the t-DFS scheme. According to the QSE program, the basic vectors of t-DFS can be used to determine the incoherent control program, and the coherent control project is also fixed by Eq. (\ref{relation}). Therefore, by elegantly combining two projects together, the quantum state will evolve following the t-DFS strictly.}\label{illu}
\end{figure}

Since the goal of the QSE program is to design a path in the Hilbert space to connect the initial state and the target state, the t-DFS is the best candidate for implementing a fast and robust QSE program. In the rest of this section, we show how to engineer a state of open quantum system to the target state. The design process is illustrated in FIG. \ref{illu}. From the condition of t-DFS, one can conclude that the t-DFS can be constructed by combining the incoherence control project with the coherent control project on the open quantum system. The two sorts of controls perform different duties. For the incoherent control, since the basic vectors of t-DFS are common eigenvectors of the Lindblad operators, the design is used to obtain time-dependent Lindblad operators whose common eigenvectors must connect the initial state with the target state. At the same time, the evolution of the quantum state must follow the t-DFS strictly, which is the duty of the coherent control part. The coherent control field is not only determined by the incoherent control design, but also restricted by the condition mentioned above, that the t-DFS must be invariant under the operator $H_{\text{eff}}$, i.e., $\bra{\Phi^\bot_i(t)} H_{\text{eff}} \ket{\Phi_j(t)}=0$ for $\forall i, j$, where $\ket{\Phi^\bot_i(t)}$, is one of the basic vectors of the componental subspace of $\mathcal H_{\text{DFS}}(t)$. Considering the concrete structure of $H_{\text{eff}} $ in Eq. (\ref{heff}), the condition mentioned above can be reduced to the following form:
\begin{eqnarray}
&&\bra{\Phi_k(t)}{H}(t)\ket{\Phi^\bot_n(t)} =-i\bra{\dot{\Phi}_k(t)}
\Phi_n^\bot(t)\rangle\nonumber\\
&&-\frac{i}{2}\sum_\alpha\gamma_\alpha
c^{\ast}_\alpha(t)\bra{\Phi_k(t)}F_\alpha(t)\ket{\Phi_n^\bot(t)}.\label{relation}
\end{eqnarray}%
As shown in the above equation, the coherent control project (the left terms) and the incoherent control project (the right terms) restrict each other. When the design on the incoherent control project is confirmed, the basic vectors of the t-DFS are determined at the same time, which also fixes the coherent control project via Eq. (\ref{relation}). On the other hand, any requirement on the coherent control field (e.g. the shape of the laser field) also limits the incoherent control project. Thus if both the coherent control and the incoherent control project are manipulated synchronously, the state of the open quantum system will be locked in the t-DFS. Therefore, the QSE process is protected completely by the t-DFS within an arbitrary total engineering time.

Here we should make some remarks on the t-DFS QSE program. (1) Although the t-DFS is a scheme by combining the coherent controls with the incoherent ones, as reported in Ref.\cite{nadi4}, the t-DFS QSE scheme completely protects the quantum state by the symmetry of the interaction between the open quantum system and its environment. (2) Different from traditional DFS QSE schemes, the basic vectors of the DFS are time-dependent, which helps us coherently engineer the quantum state even in a one-dimension t-DFS. (3) The total engineering time of the QSE is not dependent upon the decay rate of the open system; rather, it is determined by the incoherent control project.

\section{Engineering Quantum States by the t-DFS Scheme}\label{SNC}

In the above section, we proposed a realizable method for engineering the quantum state of a single atom by means of the t-DFS scheme. The interest in this topic is driven by fundamental connections to quantum physics, as well as by potential applications to quantum state measurements \cite{qm} and quantum computing \cite{qg}. In the following, we will show how to engineer the quantum states of a two-level atom into target states.

Consider a two-level atom with ground state $\ket{0}$ and excited state $\ket{1}$ coupled to both a broadband squeezed vacuum field and a coherent control field $\Omega(t)$. In the Markov approximation, the influence of the reservoir on the system of atoms can be described by the dynamical semigroup with the generator
\begin{eqnarray}
\mathcal L=-i[H,\cdot]+\mathcal L_D.\label{liou}
\end{eqnarray}%
In the rotating frame, the Hamiltonian of two-level atom can be written as
\begin{eqnarray}
H=\Omega(t)\ket{0}\bra{1}+h.c..\label{Ham}
\end{eqnarray}%
The dissipator caused by the coupling to the squeezed vacuum is
\begin{eqnarray}
\mathcal L_D \rho&=&\gamma \cosh^2(r)\left(\sigma_+\rho(t)\sigma_-
-\frac{1}{2}\{\sigma_+\sigma_-\rho(t)\}\right)\nonumber\\&&+\gamma
\sinh^2(r)\left(\sigma_-\rho(t)\sigma_+
-\frac{1}{2}\{\sigma_-\sigma_+\rho(t)\}\right)\nonumber\\
&&+\gamma\sinh(r)\cosh(r) \exp(-i \theta) \sigma_-\rho(t)\sigma_-\nonumber\\
&&+\gamma \sinh(r)\cosh(r) \exp(i \theta) \sigma_+\rho(t)\sigma_+,\label{maeq2}
\end{eqnarray}%
where $r$ is the squeezing parameter and $\theta$ is the squeezing phase; $\sigma_-$ ($\sigma_+$) is the lowering (raising) operator and $\gamma$ is the spontaneous decay rate. In Eq.(\ref{maeq2}), we have assumed that the vacuum squeezing field is perfect. If we redefine the decoherence operator as follows:
\begin{eqnarray}
L=\cosh(r)\exp(-i\theta/2)\sigma_-+\sinh(r)\exp(i\theta/2)\sigma_+,\label{Lindblad}
\end{eqnarray}%
the dissipator could be transformed into the Lindblad form,
\begin{eqnarray}
\mathcal L_D \rho=\gamma/2(2 L \rho L^\dag-\{L^\dag L, \rho \}).
\end{eqnarray}%

By definition, DFS is composed of states that undergo unitary evolution. Obviously, one-dimension (1D) DFS is inadequate to engineer the quantum states into the target state. But if the basic vectors of DFS depend on time, the DFS will evolve in the Hilbert space of the two-level atom. So we need to find the 1D DFS, and then let it evolve to the target state. This is the main idea of the t-DFS.

First, according to the NS condition of t-DFSs, a subspace spanned by $\mathcal H_t= \{\ket{\phi}\}$ is a decoherence-free subspace if $\ket{\phi}$ is the eigenvector of the Lindblad operator $L$. It is obvious that the Lindblad operator $L$ (Eq.(\ref{Lindblad})) gives two nonorthogonal eigenvectors,
\begin{eqnarray}
\ket{\phi_1}=(\sqrt{\sinh(r)}\exp(i\theta/2)\ket{0}+\sqrt{\cosh{(r)}}\ket{1})/p,\nonumber\\
\ket{\phi_2}=(-\sqrt{\sinh(r)}\exp(i\theta/2)\ket{0}+\sqrt{\cosh{(r)}}\ket{1})/p,
\end{eqnarray}%
with eigenvalues $\lambda_1=\sqrt{\sinh(r)\cosh(r)}$ and $\lambda_2=-\sqrt{\sinh(r)\cosh(r)}$, in which $p=\sinh(r)+\cosh(r)$ is the normalizing factor. Any of the eigenvectors can be the basic vector of subspace $\mathcal H_t$. To maintain generality, we choose $\ket{\phi_1}$ to construct the 1D subspace $\mathcal H_t$. At the same time, the basic vector of orthogonal complement space is also determined by
\begin{eqnarray}
\ket{\phi^\bot}=(\sqrt{\cosh(r)}\exp(i\theta/2)\ket{0}-\sqrt{\sinh{(r)}}\ket{1})/p.
\end{eqnarray}%
The set of bases $\{\ket{\phi_1},\ket{\phi^\bot}\}$ is a complete set of the Hilbert space of the two-level atom $\mathcal H$. Distinctly, the eigenvector $\ket{\phi_1}$ depends on the parameters of the squeezed vacuum, i.e., the squeezed parameter $r$ and the squeezed phase $\theta$. Assume that there is a time-dependent squeezed parameter $r(t)$ and a squeezed phase $\theta (t)$, both of which ought to be reasonably chosen and realizable in the laboratory.

In the following, we design an experimental process to create the 1D t-DFS. First, the two-level atom is placed in the vacuum field. When it couples to the vacuum, the two-level atom decays into the ground state $\ket{0}$. Here we consider a more realistic case of an extremely small population in the excited state. So the initial state we use here is $\ket{\varphi(0)}=\sqrt{1-o^2}\ket{0}+o\ket{1}$, where $o$ is an extremely small constant. After that, we engineer the surroundings of the two-level atom from the vacuum field to the squeezed vacuum field by means of engineering reservoir technology \cite{re1,re2}, which results in the time-dependence of the squeezed parameters. The way in which the parameters depend on time is determined by the scheme of the reservoir engineering\cite{svf1,svf2}. For simplicity, both the squeezed parameter and the squeezed phase are set to depend on time linearly.
\begin{eqnarray}
r(t)=\mu t+o,\ \theta=\nu t,
\end{eqnarray}%
where $\mu$ and $\nu$ are constants related to the concrete way of reservoir engineering. With the evolution of the squeezed field parameters, the subspace $\mathcal H_t$ is a time-dependent 1D subspace in which quantum state of a two-level atom is protected against decoherence. Thus the two-level atom is controlled to guarantee that the quantum state is bound to the subspace $\mathcal H_t$ at all times. In other words, the subspace $\mathcal H_t$ is t-DFS if and only if the two-level atom is controlled to make sure that the subspace $\mathcal H_t$ is invariant under $H_{\text{eff}}(t)$, as shown in Eq. (\ref{heff}). When the effective Hamiltonian $H_{\text{eff}}(t)$ acts on a quantum state $\ket{\phi}$ in the t-DFS $\mathcal H_t$, the quantum state $\ket{\varphi}=H_{\text{eff}}(t)\ket{\phi}$ is still within the t-DFS $\mathcal H_t$, i.e., $\bra{\phi^\bot}\varphi\rangle=0$. Taking the Hamiltonian Eq. (\ref{Ham}) into the effective Hamiltonian $H_{\text{eff}}(t)$ and considering the above requirement, we are able to find the accurate function of the coherent control field $\Omega(t)$. The real part $\Omega_R(t)$ and the imaginary part $\Omega_I(t)$ of the coherent control field ($\Omega(t)=\Omega_R(t)+i\Omega_I(t)$) can be written as
\begin{eqnarray}
\Omega_R(t)&=&-{\cos(\nu t) f_1(t)-\sin(\nu t) f_2(t)}\nonumber\\
\Omega_I(t)&=&{\sin(\nu t) f_1(t)-\cos(\nu t) f_2(t)}
\label{cont}
\end{eqnarray}%
with $f_1(t)=\nu\exp(-\mu t-o)\sqrt{\sinh(\mu t+o)\cosh(\mu t+o)}/2$ and $f_2(t)=\exp(-\mu t-o)(\mu/\sqrt{\sinh(\mu t+o)\cosh(\mu t+o)}+\gamma\sqrt{\sinh(\mu t+o)\cosh(\mu t+o)})/2$. By combining the reservoir engineering scheme Eq.(\ref{Lindblad}) with the coherence control field Eq.(\ref{cont}), the 1D t-DFS is constructed and the quantum state of two-level atom evolves from the ground state $\ket{0}$ to a superposition states $\ket{\phi}$ coherently. By the same way, the QSE with different initial state can also be engineered.

\begin{figure}
\includegraphics[scale=0.45]{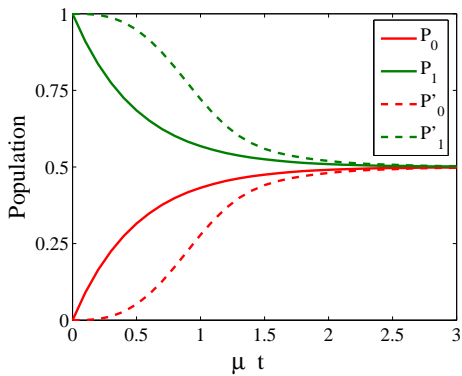}
\caption{(Color online) The population on the ground state $\ket{0}$ (red lines) and the excited state $\ket{1}$ (green lines) versus the dimensionless parameter $\mu t$. The results were obtained by calculating the master equation with the coherence control field $\Omega(t)$ (solid lines) and without the coherence control field (dash lines). The figure is evaluated for $\mu=\gamma$ and $\nu=2\pi\gamma/3$. }\label{Pt}
\end{figure}
\begin{figure}
\includegraphics[scale=0.45]{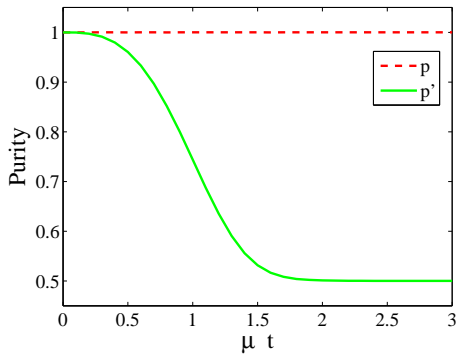}
\caption{(Color online) The purity versus the dimensionless parameter $\mu t$. The results were obtained by calculating the master equation with the coherence control field $\Omega(t)$ (red solid lines) and without the coherence control field  (green dash lines).The figure is evaluated for $\mu=\gamma$ and $\nu=2\pi\gamma/3$.}\label{purity}
\end{figure}

To judge the validity of our scheme on population engineering, we studied both the population transferring from the ground state to the excited state and the purity of the quantum state. In FIG.\ref{Pt}, the population on the ground state $\ket{0}$ (red lines) and the excited state $\ket{1}$ (green lines) are both plotted. The solid lines in the figure are the populations on the ground state ($P_0$) and the excited state ($P_1$), which are plotted according to the master equation Eq.(\ref{liou}) with the coherence control field Eq. (\ref{cont}); the dash lines are the populations on the ground state ($P'_0$) and the excited state ($P'_1$), which are plotted based on the same master equation only without the coherence control field. Here we choose $\mu=\gamma$ and $\nu=2\pi\gamma/3$. FIG. \ref{Pt} shows that either the two-level is manipulated by $\Omega(t)$ or not and that the population will definitely transfer from the ground state to the excited state; the population on the ground state and the excited state are equal when the steady state is reached.

\begin{figure}
\includegraphics[scale=0.45]{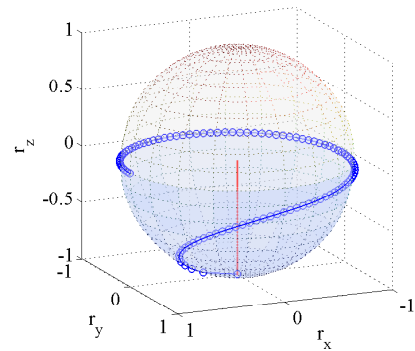}
\caption{(Color online) The evolution of the quantum state on the Bloch sphere. The results were obtained by calculating the master equation with the coherence control field $\Omega(t)$ (blue circle line) and without the coherence control field (red line). The figure is evaluated for $\mu=\gamma$ and $\nu=2\pi\gamma/3$. }\label{bloch}
\end{figure}

However, the principle behind the similarity mentioned above is different. On the one hand, when the two-level atom is not manipulated by the coherent control field Eq. (\ref{cont}), its quantum character will gradually be lost because of the coupling to the squeezed vacuum field. If we consider the purity ($p'$) of the quantum state (see the green solid line in FIG. \ref{purity}), we find it decays over time. As a consequence, the quantum state will become the maximally mixed state \cite{qo}. On the other hand, when the two-level atom is coherently controlled on the basis of Eq. (\ref{cont}), the quantum state of two-level system stabilizes in the t-DFS from the beginning to the end. The t-DFS ensures that the evolution of the quantum state is unitary and that the purity ($p$) does not change over time (see the red dash line in FIG. \ref{purity}). The results obtained above coincide with our previous prediction. We should also mention that the reason for choosing $\mu=\gamma$ is not a necessary requirement on our QSE scheme; it is used to compare our scheme with the decoherence process of a two-level atom. Theoretically speaking, the selections of $\mu$ and $\nu$ is quite arbitrary. The only limiting factor is the experimental technology on the reservoir engineering.

For illustrating the phase engineering more obviously, the Bloch vectors are plotted in FIG. \ref{bloch}. As shown by the red line in FIG. \ref{bloch}, the quantum state decays to the center of the Bloch sphere gradually in the case of absence of the coherent control field. Even though the squeezed vacuum field is engineered accordingly, there is no response of the phase to the reservoir engineering. However, when the two-level atom is manipulated by the coherent control field (Eq. (\ref{cont})), the situation is changed. The phase between the ground state and the excited state (blue circle line in FIG.\ref{bloch}) varies following our prediction, which is useful in the quantum computation program \cite{qc}.

The coherence control field is the key point for implementing the t-DFS scheme for both the population and the phase engineering. When the atom is controlled opportunely, the Bloch vector of the quantum state is on the Bloch sphere's surface. Otherwise, the Bloch vector will enter the Bloch sphere towards the zero vector \cite{qo}. Considering the asymptotic behavior of the t-DFS's basic vector,  every single point on the surface of the lower part can be reached by means of the t-DFS scheme.

\section{Adjustment of the coherent control field}\label{AD}

\begin{figure}
\includegraphics[scale=0.45]{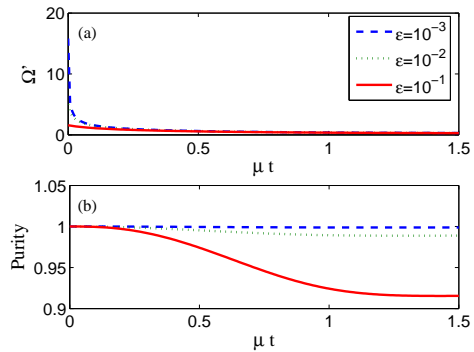}
\caption{(Color online) The evolution of (a) the control field $|\Omega'|$ and (b) the purity $p$. The results were obtained by calculating the master equation with the adjusted control function $\Omega'(\epsilon)$ with $\epsilon=10^{-3}$ (blue dash line), $\epsilon=10^{-2}$ (green dot line), and $\epsilon=10^{-1}$ (red solid line). The figure is evaluated for $\mu=\gamma$. }\label{PS}
\end{figure}

If we neglected the extremely small constant $o$ in Eq. (\ref{cont}), the coherence control field should have a singular point at $t=0$, i.e., $\lim_{o\rightarrow 0}\Omega(0)=\infty$. It is difficult to achieve such a control function experimentally. In the following, we propose another coherence control function to avoid such a case. It can be observed that the single point is caused by the denominator ($\sqrt{2\sinh(\mu t)\cosh(\mu t)}$) of the function $f_2(t)$, so that the control function of the coherent control field can be adjusted accordingly. The new control field $\Omega'(t)$ has the same structure of $\Omega(t)$ as Eq. (\ref{cont}), but the only difference is that
\begin{eqnarray}
&&f_2(t)=\frac{\exp(-\mu t)}{2}\nonumber
\\&&\times\left(\frac{\mu}{\sqrt{\sinh(\mu t+\epsilon)\cosh(\mu t)}}+\gamma\sqrt{\sinh(\mu t)\cosh(\mu t)}\right),\nonumber\\\label{cont2}
\end{eqnarray}%
where $\epsilon$ is a small constant. When we use the control function $\Omega'(t)$ instead of $\Omega(t)$, the evolution of the quantum state is not unitary and the purity must decay. Here we intend to study the effect of this modification on both the coherence control field and the purity of the quantum state. On the one hand, with the increase of constant $\epsilon$, the control field's strength $\Omega'(0)$ becomes weaker and weaker, which is advantageous in the realization of the t-DFS scheme experimentally. On the other hand, the modification of the coherence control field can not protect the quantum state perfectly. Therefore, we will concentrate on the asymptotic state of the two-level atom first. Since the adjustment on the coherent control field in such way does not affect the phase engineering and the manipulation on the phase has no effect on the asymptotic purity, it is convenient to choose $\nu=0$ in the following discussion. As a consequence, the coherent control field is given by
\begin{eqnarray}
\Omega'(t)=-i\frac{\exp(-\mu t)}{2}\frac{\left(\mu+\gamma\sinh(\mu t)\cosh(\mu t)\right)}{\sqrt{\sinh(\mu t+\epsilon)\cosh(\mu t)}}.\label{cont3}
\end{eqnarray}%
Taking the above coherent control function into Eq. (\ref{liou}), the matrix elements of the quantum state $\rho$ with respect to the basis $\ket{0}$ and $\ket{1}$ satisfy the following differential equation set:
\begin{eqnarray}
\dot \rho_{00}&=&\sinh^2(\mu t)-2i\Omega'(t)\rho_{01}-\cosh(2\mu t)\rho_{00},\nonumber\\
\dot \rho_{01}&=&i\Omega'(t)(1-2\rho_{00})-\exp(-2\mu t)\rho_{01},\label{deq}
\end{eqnarray}%
in which $\Omega'(t)=-\Omega'(t)^*$ and $\rho_{00}+\rho_{11}=1$ have been considered. Direct calculations show that the quantum state is a unique stationary asymptotic state $\rho^s$ that has non-vanishing matrix elements
\begin{eqnarray}
\rho^s_{00}=1/2,\,\, \rho^s_{01}=2 \Omega'_s,\label{matrix}
\end{eqnarray}%
with the asymptotic strength of the coherence control field $\Omega'_s=\exp(\epsilon/2)/4$. When the coherence control field is absent, the asymptotic state is the maximal mixed state; when $\epsilon=0$, the t-DFS scheme can be achieved. Generally speaking, the asymptotic purity as a function of the parameter $\epsilon$ can be written as
\begin{eqnarray}
p_s=\frac{1+\exp(-\epsilon)}{2},\label{apurity}
\end{eqnarray}%
In FIG. \ref{PS}, the evolution of both the purity and the coherence control field are plotted. With the increase of the parameter $\epsilon$, the strength of the control field is evidently reduced. At the same time, the purity also decays. However, we can see that even the control field strength is so weak (red solid line in FIG. \ref{PS}(a)) that the asymptotic purity is still high. In other words, although the coherence field cannot be reached as Eq. (\ref{cont}), the coherence of the quantum state is also robust.

\begin{figure}
\includegraphics[scale=0.45]{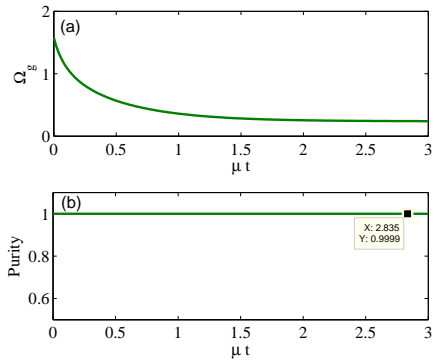}
\caption{(Color online) The evolution of (a) the control field $|\Omega_g|$ and (b) the purity $p$. The results were obtained by calculating the master equation with the control function $\Omega_g$ with $\epsilon_0=10^{-1}$ and $\Gamma=10^3\gamma$. The figure is evaluated for $\mu=\gamma$. }\label{og}
\end{figure}

What we have shown above is only an example of adjusting the control field $\Omega$, but there are still many more methods to adjust it. Generally speaking, we can introduce the adjusted control field as
\begin{eqnarray}
\Omega_g (\epsilon)=\Lambda(\epsilon)\Omega,
\end{eqnarray}%
where $\Lambda(\epsilon)$ is an adjusted function, $\epsilon$ should not have to be a constant, and $\Omega$ is the coherent control field, as shown in Eq. (\ref{cont}) with $\nu=0$.  The function $\Lambda(\epsilon)$ needs to satisfy the following conditions: 1. The adjusted control field $\Omega_g$ must be analyzed at the singular point of the control field $\Omega$; 2. Under the control of $\Omega_g$, the purity must be as robust as possible. To give an example, we introduce a simple function $\Lambda(t)=\sqrt{\sinh(\mu t)/\sinh(\mu t+\epsilon_0\exp(-\Gamma t))}$, where $\Gamma$ is the decay rate of the parameter $\epsilon_0$. Such an adjusted control field certainly satisfies both of the conditions mentioned above. For $t=0$, the singular point of the control field vanishes. And since
\begin{eqnarray}
\frac{d|\Omega_g|}{d\Gamma}&=&|\Omega|\sqrt{\frac{\sinh(\mu t+\epsilon(t))}{\sinh{\mu t}}}\nonumber\\ &\times&\frac{\Gamma\exp(-\Gamma t)\cosh(\mu t+\epsilon(t))\sinh(\mu t)}{2\sinh(\mu t+\epsilon(t))}>0,\nonumber
\end{eqnarray}%
the following inequality can be given:
\begin{eqnarray}
 \frac{d p_s}{d\Gamma}=16|\Omega_g|\frac{d |\Omega_g|}{d\Gamma}>0.\nonumber
\end{eqnarray}%
These results indicate that the more rapidly the parameter decays, the higher the purity obtained. The purity of the quantum state controlled by $\Omega'(\epsilon_0)$ is the lower limit, which is controlled by $\Omega_g$. This adjustment on the control function is so powerful that a simple and realizable control field can protect the quantum character of the two-level atom. In FIG. \ref{og}, the control field $\Omega_g$ and the purity $p$ are presented, in which the decay rate $\Gamma$ is chosen as $\Gamma=10^3\mu$ and the constant parameter is $\epsilon_0=10^{-1}$. The evolution of the quantum state is almost unitary and the purity is no less than $0.9999$. It is important to emphasize that the t-DFS scheme is universal and allows several ways of engineering the reservoir coupled to the main system. By engineering the reservoir and choosing the control field properly, the quantum state of the main system can be engineered as though the surrounding environment does not exist.

\section{Summary} \label{CC}

We have presented a proposal employing a t-DFS scheme to engineer the quantum state of an open quantum system. We showed that, although the quantum system couples to a decohering environment, the QSE process designed using our method is completely unitary within an arbitrary total engineering time. As shown in this paper, the t-DFS QSE program is designed according to an elegant combination of incoherent control and coherent control projects, which play different roles in this program.

Such a method is powerful and reliable for realizing the quantum information and the quantum simulation equipment. First, a QSE task can be implemented in a one-dimensional t-DFS with no fidelity loss. For the QSE program of the open quantum system, the previous proposals either need numerous physical qubits to construct a multi-dimensional DFS or require ultra-fast operation on single physical qubit in order to preserve the quantum characters of the open quantum system. Moreover, every single common eigenvector of the Lindblad operators with any eigenvalue can be chosen as the basic vector of the t-DFS. This provides various selections for implementing the QSE program, which is useful for finding the best scheme in realization of the QSE program. Secondly, a real quantum information process always involves numerous operations on a single qubit. A tiny loss in fidelity in one of the QSE programs will lead to the quantum information process failure after repeated operation on the same qubit. To avoid this, the QSE process must be unitary, or at least the asymptotic purity of the quantum state must be robust. Our method can achieve a unitary operation on the qubit to implement a multi-task QSE program with no loss of fidelity. Even if the effect of decoherence excites, the asymptotic purity remains satisfactory. So the proposed scheme is not only a reliable QSE process experimentally, it is also the best choice when the goal is to construct a real quantum computer or quantum communication equipment.

We thank X. L. Huang and W. B. Yan for  valuable discussions on this manuscript. This work is supported by the NSF of China under Grants  No 11347169.

\end{document}